\documentclass[reprint,amsmath,amssymb,aps,prl]{revtex4-1}
\usepackage{graphicx}
\usepackage{dcolumn}
\usepackage{bm}
\begin{document}
\title{Experimental quantum teleportation}
\author{Dik Bouwmeester}
\affiliation{Institut f\"ur Experimentalphysik, Universit\"at Innsbruck, Technikerstr. 25, A-6020 Innsbruck, Austria}
\author{Jian-Wei Pan}
\affiliation{Institut f\"ur Experimentalphysik, Universit\"at Innsbruck, Technikerstr. 25, A-6020 Innsbruck, Austria}
\author{Klaus Mattle}
\affiliation{Institut f\"ur Experimentalphysik, Universit\"at Innsbruck, Technikerstr. 25, A-6020 Innsbruck, Austria}
\author{Manfred Eibl}
\affiliation{Institut f\"ur Experimentalphysik, Universit\"at Innsbruck, Technikerstr. 25, A-6020 Innsbruck, Austria}
\author{Harald Weinfurter}
\affiliation{Institut f\"ur Experimentalphysik, Universit\"at Innsbruck, Technikerstr. 25, A-6020 Innsbruck, Austria}
\author{Anton Zeilinger}
\affiliation{Institut f\"ur Experimentalphysik, Universit\"at Innsbruck, Technikerstr. 25, A-6020 Innsbruck, Austria}
\date{December 11, 1997}

\begin{abstract}
Quantum teleportation -- the transmission and reconstruction over arbitrary distances of the state of a quantum system -- is demonstrated experimentally. During teleportation, an initial photon which carries the polarization that is to be transferred and one of a pair of entangled photons are subjected to a measurement such that the second photon of the entangled pair acquires the polarization of the initial photon. This latter photon can be arbitrarily far away from the initial one. Quantum teleportation will be a critical ingredient for quantum computation networks.
\end{abstract}
\maketitle

The dream of teleportation is to be able to travel by simply reappearing at some distant location. An object to be teleported can be fully characterized by its properties, which in classical physics can be determined by measurement. To make a copy of that object at a distant location one does not need the original parts and pieces -- all that is needed is to send the scanned information so that it can be used for reconstructing the object. But how precisely can this be a true copy of the original? What if these parts and pieces are electrons, atoms and molecules? What happens to their individual quantum properties, which according to the Heisenberg's uncertainty principle cannot be measured with arbitrary precision?

Bennett \textit{et al.}~\cite{ref:1} have suggested that it is possible to transfer the quantum state of a particle onto another particle -- the process of quantum teleportation -- provided one does not get any information about the state in the course of this transformation. This requirement can be fulfilled by using entanglement, the essential feature of quantum mechanics~\cite{ref:2-1,*ref:2-2,*ref:2-3}. It describes correlations between quantum systems much stronger than any classical correlation could be.

The possibility of transferring quantum information is one of the cornerstones of the emerging field of quantum communication and quantum computation~\cite{ref:3}. Although there is fast progress in the theoretical description of quantum information processing, the difficulties in handling quantum systems have not allowed an equal advance in the experimental realization of the new proposals. Besides the promising developments of quantum cryptography~\cite{ref:4} (the first provably secure way to send secret messages), we have only recently succeeded in demonstrating the possibility of quantum dense coding~\cite{ref:5}, a way to quantum mechanically enhance data compression. The main reason for this slow experimental progress is that, although there exist methods to produce pairs of entangled photons~\cite{ref:6}, entanglement has been demonstrated for atoms only very recently~\cite{ref:7} and it has not been possible thus far to produce entangled states of more than two quanta.

Here we report the first experimental verification of quantum teleportation. By producing pairs of entangled photons by the process of parametric down-conversion and using two-photon interferometry for analysing entanglement, we could transfer a quantum property (in our case the polarization state) from one photon to another. The methods developed for this experiment will be of great importance both for exploring the field of quantum communication and for future experiments on the foundations of quantum mechanics.

\section{The problem}
To make the problem of transferring quantum information clearer, suppose that Alice has some particle in a certain quantum state $|\psi\rangle$ and she wants Bob, at a distant location, to have a particle in that state. There is certainly the possibility of sending Bob the particle directly. But suppose that the communication channel between Alice and Bob is not good enough to preserve the necessary quantum coherence or suppose that this would take too much time, which could easily be the case if $|\psi\rangle$ is the state of a more complicated or massive object. Then, what strategy can Alice and Bob pursue?

As mentioned above, no measurement that Alice can perform on $|\psi\rangle$ will be sufficient for Bob to reconstruct the state because the state of a quantum system cannot be fully determined by measurements. Quantum systems are so evasive because they can be in a superposition of several states at the same time. A measurement on the quantum system will force it into only one of these states -- this is often referred to as the projection postulate. We can illustrate this important quantum feature by taking a single photon, which can be horizontally or vertically polarized, indicated by the states $|\leftrightarrow\rangle$ and $|\updownarrow\rangle$. It can even be polarized in the general superposition of these two states
\begin{equation}
|\psi\rangle=\alpha|\leftrightarrow\rangle+\beta|\updownarrow\rangle
\label{eq:1}
\end{equation}
where $\alpha$ and $\beta$ are two complex numbers satisfying $|\alpha|^{2}+|\beta|^{2}=1$. To place this example in a more general setting we can replace the states $|\leftrightarrow\rangle$ and $|\updownarrow\rangle$ in Eq.~(\ref{eq:1}) by $|0\rangle$ and $|1\rangle$, which refer to the states of any two-state quantum system. Superpositions of $|0\rangle$ and $|1\rangle$ are called qubits to signify the new possibilities introduced by quantum physics into information science~\cite{ref:8}.

If a photon in state $|\psi\rangle$ passes through a polarizing beamsplitter -- a device that reflects (transmits) horizontally (vertically) polarized photons -- it will be found in the reflected (transmitted) beam with probability $|\alpha|^{2}$ ($|\beta|^{2}$). Then the general state $|\psi\rangle$ has been projected either onto $|\leftrightarrow\rangle$ or onto $|\updownarrow\rangle$ by the action of the measurement. We conclude that the rules of quantum mechanics, in particular the projection postulate, make it impossible for Alice to perform a measurement on $|\psi\rangle$ by which she would obtain all the information necessary to reconstruct the state.

\section{The concept of quantum teleportation}
Although the projection postulate in quantum mechanics seems to bring Alice's attempts to provide Bob with the state $|\psi\rangle$ to a halt, it was realised by Bennett et al.~\cite{ref:1} that precisely this projection postulate enables teleportation of $|\psi\rangle$ from Alice to Bob. During teleportation Alice will destroy the quantum state at hand while Bob receives the quantum state, with neither Alice nor Bob obtaining information about the state $|\psi\rangle$. A key role in the teleportation scheme is played by an entangled ancillary pair of particles which will be initially shared by Alice and Bob.

\begin{figure}
\includegraphics[width=3.4in]{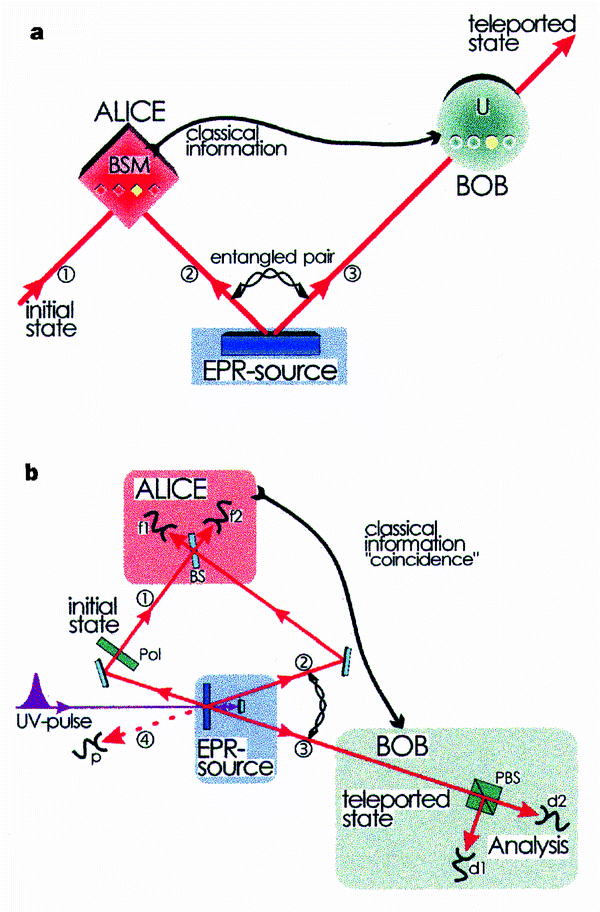}
\caption{Scheme showing principles involved in quantum teleportation (a) and the experimental set-up (b). (a) Alice has a quantum system, particle 1, in an initial state which she wants to teleport to Bob. Alice and Bob also share an ancillary entangled pair of particles 2 and 3 emitted by an Einstein-Podolsky-Rosen (EPR) source. Alice then performs a joint Bell-state measurement (BSM) on the initial particle and one of the ancillaries, projecting them also onto an entangled state. After she has sent the result of her measurement as classical information to Bob, he can perform a unitary transformation (U) on the other ancillary particle resulting in it being in the state of the original particle. (b) A pulse of ultraviolet radiation passing through a nonlinear crystal creates the ancillary pair of photons 2 and 3. After retroflection during its second passage through the crystal the ultraviolet pulse creates another pair of photons, one of which will be prepared in the initial state of photon 1 to be teleported, the other one serving as a trigger indicating that a photon to be teleported is under way. Alice then looks for coincidences after a beam splitter BS where the initial photon and one of the ancillaries are superposed. Bob, after receiving the classical information that Alice obtained a coincidence count in detectors f1 and f2 identifying the $|\psi^{-}\rangle_{12}$ Bell state, knows that his photon 3 is in the initial state of photon 1 which he then can check using polarization analysis with the polarizing beam splitter PBS and the detectors d1 and d2. The detector p provides the information that photon 1 is under way.}
\label{fig:1}
\end{figure}

Suppose particle 1 which Alice wants to teleport is in the initial state $|\psi\rangle_{1}=\alpha|\leftrightarrow\rangle_{1}+\beta|\updownarrow\rangle_{1}$ (Fig.~\ref{fig:1}(a)), and the entangled pair of particles 2 and 3 shared by Alice and Bob is in the state:
\begin{equation}
|\psi^{-}\rangle_{23}=\frac{1}{\quad\overline{2}}|\leftrightarrow\rangle_{2}|\updownarrow\rangle_{3}-|\updownarrow\rangle_{2}|\leftrightarrow\rangle_{3}
\label{eq:2}
\end{equation}

That entangled pair is a single quantum system in an equal superposition of the states ${|\leftrightarrow\rangle_{2}|\updownarrow\rangle_{3}}$ and ${|\updownarrow\rangle_{2}|\leftrightarrow\rangle_{3}}$. The entangled state contains no information on the individual particles; it only indicates that the two particles will be in opposite states. The important property of an entangled pair is that as soon as a measurement on one of the particles projects it, say, onto $|\leftrightarrow\rangle$ the state of the other one is determined to be $|\updownarrow\rangle$, and vice versa. How could a measurement on one of the particles instantaneously influence the state of the other particle, which can be arbitrarily far away? Einstein, among many other distinguished physicists, could simply not accept this ``spooky action at a distance''. But this property of entangled states has now been demonstrated by numerous experiments (for reviews, see refs.~\onlinecite{ref:9,ref:10}).

The teleportation scheme works as follows. Alice has the particle 1 in the initial state $|\psi\rangle_{1}$ and particle 2. Particle 2 is entangled with particle 3 in the hands of Bob. The essential point is to perform a specific measurement on particles 1 and 2 which projects them onto the entangled state:
\begin{equation}
|\psi^{-}\rangle_{12}=\frac{1}{\quad\overline{2}}|\leftrightarrow\rangle_{1}|\updownarrow\rangle_{2}-|\updownarrow\rangle_{1}|\leftrightarrow\rangle_{2}
\label{eq:3}
\end{equation}

This is only one of four possible maximally entangled states into which any state of two particles can be decomposed. The projection of an arbitrary state of two particles onto the basis of the four states is called a Bell-state measurement. The state given in Eq.~(\ref{eq:3}) distinguishes itself from the three other maximally entangled states by the fact that it changes sign upon interchanging particle 1 and particle 2. This unique antisymmetric feature of $|\psi^{-}\rangle_{12}$ will play an important role in the experimental identification, that is, in measurements of this state.

Quantum physics predicts~\cite{ref:1} that once particles 1 and 2 are projected into $|\psi^{-}\rangle_{12}$, particle 3 is instantaneously projected into the initial state of particle 1. The reason for this is as follows. Because we observe particles 1 and 2 in the state $|\psi^{-}\rangle_{12}$ we know that whatever the state of particle 1 is, particle 2 must be in the opposite state, that is, in the state orthogonal to the state of particle 1. But we had initially prepared particle 2 and 3 in the state $|\psi^{-}\rangle_{23}$, which means that particle 2 is also orthogonal to particle 3. This is only possible if particle 3 is in the same state as particle 1 was initially. The final state of particle 3 is therefore:
\begin{equation}
|\psi\rangle_{3}=\alpha|\leftrightarrow\rangle_{3}+\beta|\updownarrow\rangle_{3}
\label{eq:4}
\end{equation}
We note that during the Bell-state measurement particle 1 loses its identity because it becomes entangled with particle 2. Therefore the state $|\psi\rangle_{1}$ is destroyed on Alice's side during teleportation.

This result (Eq.~(\ref{eq:4})) deserves some further comments. The transfer of quantum information from particle 1 to particle 3 can happen over arbitrary distances, hence the name teleportation. Experimentally, quantum entanglement has been shown~\cite{ref:11} to survive over distances of the order of 10 km. We note that in the teleportation scheme it is not necessary for Alice to know where Bob is. Furthermore, the initial state of particle 1 can be completely unknown not only to Alice but to anyone. It could even be quantum mechanically completely undefined at the time the Bell-state measurement takes place. This is the case when, as already remarked by Bennett \textit{et al.}~\cite{ref:1}, particle 1 itself is a member of an entangled pair and therefore has no well-defined properties on its own. This ultimately leads to entanglement swapping~\cite{ref:12,ref:13}.

It is also important to notice that the Bell-state measurement does not reveal any information on the properties of any of the particles. This is the very reason why quantum teleportation using coherent two-particle superpositions works, while any measurement on one-particle superpositions would fail. The fact that no information whatsoever is gained on either particle is also the reason why quantum teleportation escapes the verdict of the no-cloning theorem~\cite{ref:14}. After successful teleportation particle 1 is not available in its original state any more, and therefore particle 3 is not a clone but is really the result of teleportation.

A complete Bell-state measurement can not only give the result that the two particles 1 and 2 are in the antisymmetric state, but with equal probabilities of 25\% we could find them in any one of the three other entangled states. When this happens, particle 3 is left in one of three different states. It can then be brought by Bob into the original state of particle 1 by an accordingly chosen transformation, independent of the state of particle 1, after receiving via a classical communication channel the information on which of the Bell-state results was obtained by Alice. Yet we note, with emphasis, that even if we chose to identify only one of the four Bell states as discussed above, teleportation is successfully achieved, albeit only in a quarter of the cases.

\section{Experimental realization}
Teleportation necessitates both production and measurement of entangled states; these are the two most challenging tasks for any experimental realization. Thus far there are only a few experimental techniques by which one can prepare entangled states, and there exist no experimentally realized procedures to identify all four Bell states for any kind of quantum system. However, entangled pairs of photons can readily be generated and they can be projected onto at least two of the four Bell states.

\begin{figure}
\includegraphics[width=1.7in]{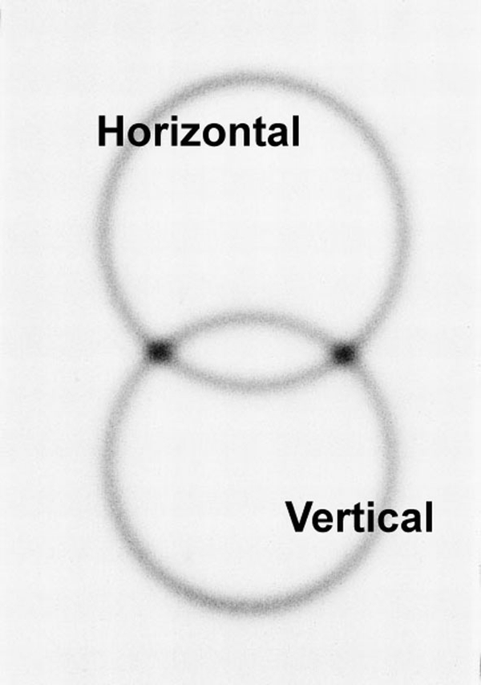}
\caption{Photons emerging from type II down-conversion (see text). Photograph taken perpendicular to the propagation direction. Photons are produced in pairs. A photon on the top circle is horizontally polarized while its exactly opposite partner in the bottom circle is vertically polarized. At the intersection points their polarizations are undefined; all that is known is that they have to be different, which results in entanglement.}
\label{fig:2}
\end{figure}

We produced the entangled photons 2 and 3 by parametric down-conversion. In this technique, inside a nonlinear crystal, an incoming pump photon can decay spontaneously into two photons which, in the case of type II parametric down-conversion, are in the state given by Eq.~(\ref{eq:2}) (Fig.~\ref{fig:2})~\cite{ref:6}.

To achieve projection of photons 1 and 2 into a Bell state we have to make them indistinguishable. To achieve this indistinguishability we superpose the two photons at a beam splitter (Fig.~\ref{fig:1}(b)). Then if they are incident one from each side, how can it happen that they emerge still one on each side? Clearly this can happen if they are either both reflected or both transmitted. In quantum physics we have to superimpose the amplitudes for these two possibilities. Unitarity implies that the amplitude for both photons being reflected obtains an additional minus sign. Therefore, it seems that the two processes cancel each other. This is, however, only true for a symmetric input state. For an antisymmetric state, the two possibilities obtain another relative minus sign, and therefore they constructively interfere~\cite{ref:15,ref:16}. It is thus sufficient for projecting photons 1 and 2 onto the antisymmetric state $|\psi^{-}\rangle_{12}$ to place detectors in each of the outputs of the beam splitter and to register simultaneous detections (coincidence)~\cite{ref:17,ref:18,ref:19}.

To make sure that photons 1 and 2 cannot be distinguished by their arrival times, they were generated using a pulsed pump beam and sent through narrow-bandwidth filters producing a coherence time much longer than the pump pulse length~\cite{ref:20}. In the experiment, the pump pulses had a duration of 200 fs at a repetition rate of 76~MHz. Observing the down-converted photons at a wavelength of 788 nm and a bandwidth of 4 nm results in a coherence time of 520 fs. It should be mentioned that, because photon 1 is also produced as part of an entangled pair, its partner can serve to indicate that it was emitted.

How can one experimentally prove that an unknown quantum state can be teleported? First, one has to show that teleportation works for a (complete) basis, a set of known states into which any other state can be decomposed. A basis for polarization states has just two components, and in principle we could choose as the basis horizontal and vertical polarization as emitted by the source. Yet this would not demonstrate that teleportation works for any general superposition, because these two directions are preferred directions in our experiment. Therefore, in the first demonstration we choose as the basis for teleportation the two states linearly polarized at $-45^{\circ}$ and $+45^{\circ}$ which are already superpositions of the horizontal and vertical polarizations. Second, one has to show that teleportation works for superpositions of these base states. Therefore we also demonstrate teleportation for circular polarization.

\section{Results}
In the first experiment photon 1 is polarized at $45^{\circ}$. Teleportation should work as soon as photon 1 and 2 are detected in the $|\psi^{-}\rangle_{12}$ state, which occurs in 25\% of all possible cases. The $|\psi^{-}\rangle_{12}$ state is identified by recording a coincidence between two detectors, f1 and f2, placed behind the beam splitter (Fig.~\ref{fig:1}(b)).

If we detect a f1f2 coincidence (between detectors f1 and f2), then photon 3 should also be polarized at $45^{\circ}$. The polarization of photon 3 is analysed by passing it through a polarizing beam splitter selecting $+45^{\circ}$ and $-45^{\circ}$ polarization. To demonstrate teleportation, only detector d2 at the $+45^{\circ}$ output of the polarizing beam splitter should click (that is, register a detection) once detectors f1 and f2 click. Detector d1 at the $-45^{\circ}$ output of the polarizing beam splitter should not detect a photon. Therefore, recording a three-fold coincidence d2f1f2 ($+45^{\circ}$ analysis) together with the absence of a three-fold coincidence d1f1f2 ($-45^{\circ}$ analysis) is a proof that the polarization of photon 1 has been teleported to photon~3.

To meet the condition of temporal overlap, we change in small steps the arrival time of photon 2 by changing the delay between the first and second down-conversion by translating the retroflection mirror (Fig.~\ref{fig:1}(b)). In this way we scan into the region of temporal overlap at the beam splitter so that teleportation should occur.

\begin{figure}
\includegraphics[width=1.7in]{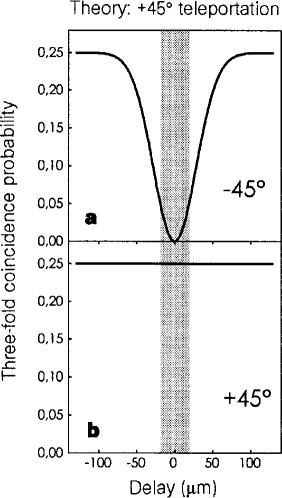}
\caption{Theoretical prediction for the three-fold coincidence probability between the two Bell-state detectors (f1, f2) and one of the detectors analysing the teleported state. The signature of teleportation of a photon polarization state at $+45^{\circ}$ is a dip to zero at zero delay in the three-fold coincidence rate with the detector analysing $-45^{\circ}$ (d1f1f2) (a) and a constant value for the detector analysis $+45^{\circ}$ (d2f1f2) (b). The shaded area indicates the region of teleportation.}
\label{fig:3}
\end{figure}

Outside the region of teleportation, photon 1 and 2 each will go either to f1 or to f2 independent of one another. The probability of having a coincidence between f1 and f2 is therefore 50\%, which is twice as high as inside the region of teleportation. Photon 3 should not have a well-defined polarization because it is part of an entangled pair. Therefore, d1 and d2 have both a 50\% chance of receiving photon 3. This simple argument yields a 25\% probability both for the $-45^{\circ}$ analysis (d1f1f2 coincidences) and for the $+45^{\circ}$ analysis (d2f1f2 coincidences) outside the region of teleportation. Figure 3 summarizes the predictions as a function of the delay. Successful teleportation of the $+45^{\circ}$ polarization state is then characterized by a decrease to zero in the $-45^{\circ}$ analysis (Fig.~\ref{fig:3}(a)), and by a constant value for the $+45^{\circ}$ analysis (Fig.~\ref{fig:3}(b)).

The theoretical prediction of Fig.~\ref{fig:3} may easily be understood by realizing that at zero delay there is a decrease to half in the coincidence rate for the two detectors of the Bell-state analyser, f1 and f2, compared with outside the region of teleportation. Therefore, if the polarization of photon 3 were completely uncorrelated to the others the three-fold coincidence should also show this dip to half. That the right state is teleported is indicated by the fact that the dip goes to zero in Fig.~\ref{fig:3}(a) and that it is filled to a flat curve in Fig.~\ref{fig:3}(b).

We note that equally as likely as the production of photons 1, 2 and 3 is the emission of two pairs of down-converted photons by a single source. Although there is no photon coming from the first source (photon 1 is absent), there will still be a significant contribution to the three-fold coincidence rates. These coincidences have nothing to do with teleportation and can be identified by blocking the path of photon 1.

\begin{figure}
\includegraphics[width=3.4in]{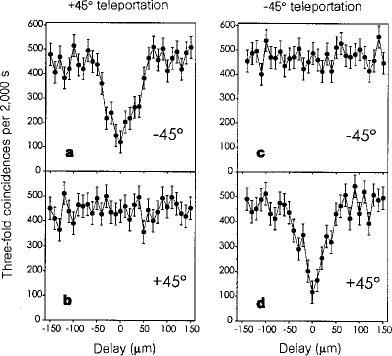}
\caption{Experimental results. Measured three-fold coincidence rates d1f1f2 ($-45^{\circ}$) and d2f1f2 ($+45^{\circ}$) in the case that the photon state to be teleported is polarized at $+45^{\circ}$ ((a) and (b)) or at $-45^{\circ}$ ((c) and (d)). The coincidence rates are plotted as function of the delay between the arrival of photon 1 and 2 at Alice's beam splitter (see Fig.~\ref{fig:1}(b)). The three-fold coincidence rates are plotted after subtracting the spurious three-fold contribution (see text). These data, compared with Fig.~{fig:3}, together with similar ones for other polarizations (Table~\ref{tab:1}) confirm teleportation for an arbitrary state.}
\label{fig:4}
\end{figure}

The probability for this process to yield spurious two- and three-fold coincidences can be estimated by taking into account the experimental parameters. The experimentally determined value for the percentage of spurious three-fold coincidences is $68\%\pm1\%$. In the experimental graphs of Fig.~\ref{fig:4} we have subtracted the experimentally determined spurious coincidences.

The experimental results for teleportation of photons polarized under $+45^{\circ}$ are shown in the left-hand column of Fig.~\ref{fig:4}; Fig.~\ref{fig:4}(a) and (b) should be compared with the theoretical predictions shown in Fig.~\ref{fig:3}. The strong decrease in the $-45^{\circ}$ analysis, and the constant signal for the $+45^{\circ}$ analysis, indicate that photon 3 is polarized along the direction of photon 1, confirming teleportation.

The results for photon 1 polarized at $-45^{\circ}$ demonstrate that teleportation works for a complete basis for polarization states (right-hand column of Fig.~\ref{fig:4}). To rule out any classical explanation for the experimental results, we have produced further confirmation that our procedure works by additional experiments. In these experiments we teleported photons linearly polarized at $0^{\circ}$ and at $90^{\circ}$, and also teleported circularly polarized photons. The experimental results are summarized in Table~\ref{tab:1}, where we list the visibility of the dip in three-fold coincidences, which occurs for analysis orthogonal to the input polarization.

\begin{table}
\caption{Visibility of teleportation in three-fold coincidences}
\label{tab:1}
\begin{ruledtabular}
\begin{tabular}{lcc}
Polarization&Visibility\\
\hline
$+45^{\circ}$&$0.63\pm0.02$&\\
$-45^{\circ}$&$0.64\pm0.02$&\\
$0^{\circ}$&$0.66\pm0.02$&\\
$90^{\circ}$&$0.61\pm0.02$&\\
Circular&$0.57\pm0.02$&\\
\end{tabular}
\end{ruledtabular}
\end{table}

\begin{figure}
\includegraphics[width=3.4in]{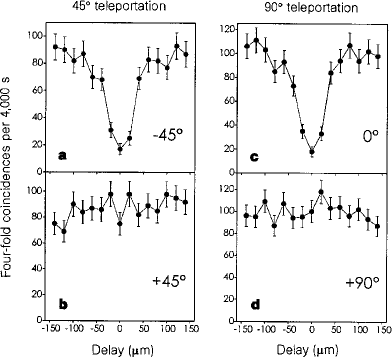}
\caption{Four-fold coincidence rates (without background subtraction). Conditioning the three-fold coincidences as shown in Fig.~\ref{fig:4} on the registration of photon 4 (see Fig.~\ref{fig:1}(b)) eliminates the spurious three-fold background. a and b show the four-fold coincidence measurements for the case of teleportation of the $+45^{\circ}$ polarization state; c and d show the results for the $+90^{\circ}$ polarization state. The visibilities, and thus the polarizations of the teleported photons, obtained without any background subtraction are $70\%\pm3\%$. These results for teleportation of two non-orthogonal states prove that we have demonstrated teleportation of the quantum state of a single photon.}
\label{fig:5}
\end{figure}

As mentioned above, the values for the visibilities are obtained after subtracting the offset caused by spurious three-fold coincidences. These can experimentally be excluded by conditioning the three-fold coincidences on the detection of photon 4, which effectively projects photon 1 into a single-particle state. We have performed this four-fold coincidence measurement for the case of teleportation of the $+45^{\circ}$ and $+90^{\circ}$ polarization states, that is, for two non-orthogonal states. The experimental results are shown in Fig.~\ref{fig:5}. Visibilities of $70\%\pm3\%$ are obtained for the dips in the orthogonal polarization states. Here, these visibilities are directly the degree of polarization of the teleported photon in the right state. This proves that we have demonstrated teleportation of the quantum state of a single photon.

\section{The next steps}
In our experiment, we used pairs of polarization entangled photons as produced by pulsed down-conversion and two-photon interferometric methods to transfer the polarization state of one photon onto another one. But teleportation is by no means restricted to this system. In addition to pairs of entangled photons or entangled atoms~\cite{ref:7, ref:21}, one could imagine entangling photons with atoms, or phonons with ions, and so on. Then teleportation would allow us to transfer the state of, for example, fast-decohering, short-lived particles, onto some more stable systems. This opens the possibility of quantum memories, where the information of incoming photons is stored on trapped ions, carefully shielded from the environment.

Furthermore, by using entanglement purification~\cite{ref:22} -- a scheme of improving the quality of entanglement if it was degraded by decoherence during storage or transmission of the particles over noisy channels -- it becomes possible to teleport the quantum state of a particle to some place, even if the available quantum channels are of very poor quality and thus sending the particle itself would very probably destroy the fragile quantum state. The feasibility of preserving quantum states in a hostile environment will have great advantages in the realm of quantum computation. The teleportation scheme could also be used to provide links between quantum computers.

Quantum teleportation is not only an important ingredient in quantum information tasks; it also allows new types of experiments and investigations of the foundations of quantum mechanics. As any arbitrary state can be teleported, so can the fully undetermined state of a particle which is member of an entangled pair. Doing so, one transfers the entanglement between particles. This allows us not only to chain the transmission of quantum states over distances, where decoherence would have already destroyed the state completely, but it also enables us to perform a test of Bell's theorem on particles which do not share any common past, a new step in the investigation of the features of quantum mechanics. Last but not least, the discussion about the local realistic character of nature could be settled firmly if one used features of the experiment presented here to generate entanglement between more than two spatially separated particles~\cite{ref:23, ref:24}.

\begin{acknowledgements}
We thank C. Bennett, I. Cirac, J. Rarity, W. Wootters and P. Zoller for discussions, and M. Zukowski for suggestions about various aspects of the experiments. This work was supported by the Austrian Science Foundation FWF, the Austrian Academy of Sciences, the TMR program of the European Union and the US NSF.
\end{acknowledgements}

\bibliography{teleportation}
\end{document}